\documentclass[12pt]{article}
\usepackage{epsfig}
\newcommand{\bkt}[1]{\langle #1 \rangle_{T}}
\newcommand{\lesssim}
{\mathrel{\raisebox{-2.8pt}{\mbox{$\stackrel{\textstyle <}{\sim}$}}}}
\newcommand{\gtrsim}
{\mathrel{\raisebox{-2.8pt}{\mbox{$\stackrel{\textstyle >}{\sim}$}}}}

\begin{document}
\begin{center}
{\bf Restoration of Isotropy in the Ising Model on the Sierpi\'nski Gasket}

\par\bigskip
Naoto Yajima\footnote{
{\em present address:}\/ 
JAMCO cooperation, 1-100 Takamatsu-cho, Tachikawa-shi, Tokyo 190-0011, Japan.
}

\par\
{\footnotesize\sl Graduate School of Human and Environmental Studies, Kyoto University, Kyoto 606-8501, Japan}
\end{center}

\begin{abstract}
We study the ferromagnetic Ising model on the Sierpi\'nski gasket (SG),
where
 the spin-spin interactions depends on the direction. Using the
 renormalization group method, we show that the ratios of the correlation
 lengths restore the isotropy of the lattice as the temperature
 approaches zero. 
 This restoration is either partial or perfect, depending
 on the interactions. In case of  partial restoration, we also evaluate
 the leading-order singular behavior of the correlation lengths.
\end{abstract}

\section{Introduction}
Restoration of microscopic isotropy in fractals with microscopic
anisotropy is observed in various physical systems, including
percolation \cite{VK}, 
diffusion and random walks \cite{wat,barlow}, resister
networks\cite{mkth}, and Gaussian field theories \cite{hat}. 
Such
phenomena are unique 
in the sense that they are absent in uniform media, while they are universal in the sense that they are
observed in a wide class of fractals. 

The isotropic spin 1/2 Ising model on the Sierpi\'nski gasket (SG) is known to have no phase transitions at finite temperature.
But the correlation length exhibits a drastic divergence as the temperature approaches zero \cite{NF,GMA}.
It would be quite interesting to determine whether the anisotropic version of the model exhibits isotropy restoration.

Indeed, Brody and Ritz \cite{BR}, and later Bhattacharyya and Chakrabarti \cite{BC} studied this problem in a partially anisotropic Ising model  (which is obtained by setting $J_2=J_3$ or $J_1=J_2$ in our model) on the SG.
Interestingly, they reached  opposite conclusions: Ref.  \cite{BR} concluded that there is a complete restoration, while  Ref. \cite{BC} concluded that restoration is impossible.

In the present paper, we carefully examine this (controversial)  problem of isotropy restoration in the Ising model on the SG.
We study a general anisotropic model in which spin-spin interactions take  three different values, according
to the directions of the bonds. 

In order to test for the possible restoration of isotropy, we define the
correlation length for each of the three directions of the
lattice. When the temperature is sufficiently high, the ratios among
the three correlation lengths are almost the same as those of the
interactions. We conclude that the behavior of the correlation lengths
respects the anisotropy of the system.

A nontrivial question is whether the isotropy is restored when the
temperature approaches zero and the correlation lengths diverge. We have
examined the behavior of the correlation lengths  using the
renormalization group, and found that the isotropy is indeed
restored. However, we also found that, depending on the interactions, the
restoration can be either perfect or partial. When  perfect restoration
takes place, all the ratios among the correlation lengths approach unity
as the temperature approaches zero. When a partial restoration takes
place, on the other hand, the ratio of the two correlation lengths
approaches unity, and the other correlation length becomes infinitely
larger, as the temperature approaches zero. 

It should be stressed that our result agrees with neither the conclusion of Ref. \cite{BR} nor that of Ref. \cite{BC}.
We have shown that the actual phenomena are much richer than those predicted in Refs. \cite{BR} or \cite{BC}.
We do, however, understand why such different conclusions have been reached.
As we  discuss  in Section~\ref{diff}, both Refs. \cite{BR} and \cite{BC} essentially rely on  simple approximate forms of the RG equation.
But, as will be clear from the careful study presented here, the RG flow in this problem cannot be reduced to a single approximate map; one needs to use different forms in different regions of the parameter space.
We thus believe that our analysis, which is clearly more elaborate and careful than the previous two, resolves the controversy and reveals the true behavior of the model.

We also stress that no distinction between
 perfect and partial restoration of isotropy, as we have found in the present model, has  been observed
in other models \cite{VK,wat,barlow,mkth,hat} in which only perfect
restoration was found. 
We still do not know why  partial
restoration is found only in the Ising model.

Let us stress that the restoration of isotropy is a phenomenon peculiar
to physical systems on fractals. In the Ising model on regular lattices,
it is believed that the ratios of the correlation lengths respect the
anisotropy of the underlying interactions, no matter how large the
correlation lengths are.

\section{Model and main results}
Here we define the model and state our main results. 

\begin{figure}
\centerline{\epsfig{file=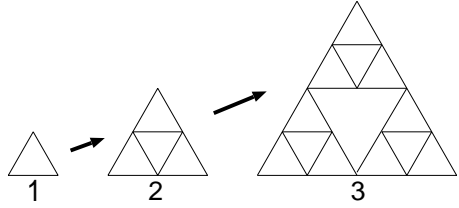,width=6cm}}
        \caption[dummy]{The first three generations of the Sierpi\'{n}ski gasket.}
        \label{fig:1}
\end{figure}

We construct the SG following the standard procedure (see Fig.~\ref{fig:1}). 
We consider an equilateral triangle formed by bonds of unit
length, and call this the first generation SG. We then put three of these
triangles together to construct the second generation SG, which is
identical to the original triangle with all sides doubled in length.
In the next step, we put three second
generation SGs together and form the third generation SG. By continuing
this procedure recursively, we  construct the $n$-th generation SG.

\begin{figure}
\centerline{\epsfig{file=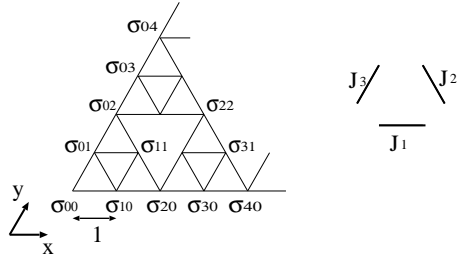,width=6cm}}
	\caption[dummy]{Ising model on the Sierpi\'{n}ski gasket. The spins are
 labeled according to the oblique coordinates.}
	\label{fig:2}
\end{figure}

We define an Ising model on the $n$-th generation SG by associating with
each site an Ising spin. The Hamiltonian of this system is
\begin{eqnarray}
 {\mathcal H}_n &=& -J_{1}( \sigma_{00}\sigma_{10}+\sigma_{10}\sigma_{20}+\sigma_{01}\sigma_{11}+\cdots ) \nonumber \\
        & &  -J_{2}( \sigma_{10}\sigma_{01}+\sigma_{20}\sigma_{11}+\sigma_{11}\sigma_{02}+\cdots ) 
        \nonumber  \\
      & & -J_{3}( \sigma_{00}\sigma_{01}+\sigma_{10}\sigma_{11}+\sigma_{01}\sigma_{02}+\cdots ),\label{eq:ham}
\end{eqnarray}
where the sums are taken over all the pairs of neighboring spins. We
label each spin variable according to the oblique coordinates, as in
Fig.~\ref{fig:2}. The positive constants $J_1$, $J_2$ and $J_3$ denote the
exchange interactions in the three directions.

The thermal expectation value of any function $A$ at  temperature $T$ is
given by
\begin{equation}
\bkt{A}^{(n)} = \frac{{\rm Tr}\,Ae^{-\beta{\mathcal H}_n}}
{{\rm Tr}\,e^{-\beta{\mathcal H}_n}},\label{eq:A}
\end{equation}
where ${\rm Tr}$ is a shorthand for the sum over all
the spins  $\sigma_{ij}=\pm1$ , and $\beta=1/k_{\rm B}T$ is the inverse temperature.

\begin{figure}
 \centerline{\epsfig{file=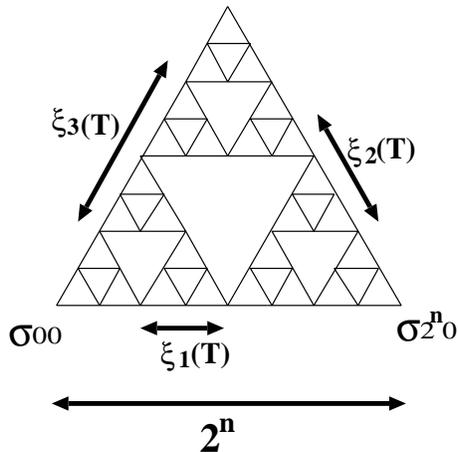,width=6cm}}
       \caption[dummy]{$\xi_1(T)$, $\xi_2 (T)$ and $\xi_3
(T)$ are the correlation lengths in the three directions.}
 	\label{fig.3}
 \end{figure}

We define the correlation length in the horizontal direction $\xi_1(T)$
by
\begin{equation}
\xi_1 (T) = \lim_{n \to \infty} \frac{2^n}{-\ln \bkt{\sigma_{00} \sigma_{2^n 0}}^{(n+1)}} . \label{eq:modoki}
\end{equation}
Similarly, we define the correlation lengths $\xi_2(T)$ and $\xi_3(T)$ for
the other two directions as in Fig.~\ref{fig.3}.

Our definition (\ref{eq:modoki}) of the correlation length is 
 motivated by the observation that the correlation between  spins at  two
 ends of the lattice should behave as
 $\bkt{\sigma_{00} \sigma_{2^n 0}}^{(n+1)}\approx\exp[-2^n/\xi_1(T)]$.
 This is apparently different from  the usual definition, motivated by the observation that
 $\bkt{\sigma_x \sigma_y}\approx\exp[-|x-y|/\xi]$ for any sites $x$, $y$.
 For the Ising model on a regular lattice, however, the two definitions are
 known to give the same correlation length (at least in the high-temperature
 region).
 Here we conjecture  equivalence, and therefor employ the simpler definition, (\ref{eq:modoki}).

\begin{figure}
\centerline{\epsfig{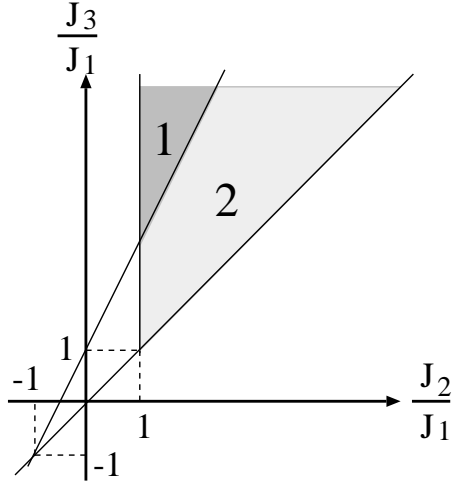}}
 	\caption[dummy]{The region 1 is where $J_3/J_1\geq(2J_2)/J_1+1$, and
 	region 2 is where $J_3/J_1<(2J_2)/J_1+1$.
 	We have partial and complete restoration of anisotropy in 
 	these two regions, respectively.}
 	\label{fig:4}
 \end{figure}

We can now state our main results. We choose the interactions $J_1$, $J_2$ and $J_3$ so as to satisfy $0<J_1\leq J_2\leq J_3$. When the temperature approaches
zero (which is the critical point of the model), the correlation lengths
diverge.  For the ratios of the correlation lengths, we show that
as $T\to0$ we have
\begin{eqnarray}
 \frac{\xi_2(T)}{\xi_1(T)} \to 1, 
 \quad
 \frac{\xi_3(T)}{\xi_1(T)} \to \infty
 \label{eq:partial}
\end{eqnarray}
if
\begin{eqnarray}
 \frac{J_3}{J_1} \geq \frac{2J_2}{J_1}+1,
  \label{eq:partial2}
\end{eqnarray}
which corresponds to the shaded region 1 in Fig.~\ref{fig:4}, and 
\begin{eqnarray}
 \frac{\xi_2(T)}{\xi_1(T)} \to 1, \quad
 \frac{\xi_3(T)}{\xi_1(T)} \to 1
  \label{eq:perfect}
\end{eqnarray}
if
\begin{eqnarray}
 \frac{J_3}{J_2}<\frac{2J_2}{J_1}+1,
  \label{eq:perfect2}
\end{eqnarray}
which corresponds to the shaded region 2 in Fig.~\ref{fig:4}. These correspond to
the partial and  perfect restoration of isotropy 
discussed in the introduction.

Moreover, when the interactions satisfy (\ref{eq:partial2}), and we have a
partial isotropy restoration (\ref{eq:partial}), we can show that the
leading-order singular behavior of the correlation lengths is given by 
\begin{eqnarray}
\xi_1(T) &\simeq& \xi_2(T) \simeq \exp\left[(\ln 2)e^{2(J_1+J_2)/(k_{B}T)}\right],\\
\xi_3(T) &\simeq& \exp\left[(\ln 2)e^{2(J_1+J_2)/(k_{B}T)}\right]\exp\left[\frac{2(J_3-J_1-2J_2)}{k_{B}T}\right]
\end{eqnarray}
as $T \to 0$. We see that the correlation lengths diverge as a
double exponential of $1/(k_{B}T)$, while the ratio $\xi_3(T)/\xi_1(T)$
diverges as a single exponential of $1/(k_{B}T)$. (The double
exponential behavior is peculiar to the Ising model on the SG, and it was previously found  in the isotropic model with $J_1=J_2=J_3$.  See Ref.  \cite{GMA}.) 

In the next section, we derive these results  using the exact
renormalization group method.

\section{Derivation}
\subsection{Renormalization group transformation}
It is well known that various models on the SG can be treated  using
the exact renormalization group transformation (usually called ``decimation''),
in which one sums up parts of the degrees of freedom.

Let ${\mathcal H}_n$ be the Hamiltonian of the form (\ref{eq:ham}) on the $n$-th
generation SG, and let $A[(\sigma_{xy})]$ be an arbitrary function of
$\sigma_{xy}$ with both $x$ and $y$  even. We denote by ${\rm Tr}^{\prime}$
the sum over $\sigma_{xy}=\pm1$ for all the sites $xy$ with both $x$ and
$y$ even. 
Let ${\rm Tr}^{\prime \prime}$ denote the sum over
$\sigma_{xy}=\pm1$ for the rest of the sites. Then from (\ref{eq:A}), we
find that
\begin{equation}
 \bkt{A}^{(n)}=\frac{{\rm Tr}Ae^{-\tilde{\mathcal H}_n}}{{\rm Tr}e^{-\tilde{\mathcal H}_n}}=\frac{{\rm Tr}^{\prime}A{\rm Tr}^{\prime \prime}e^{-\tilde{\mathcal H}_n}}{{\rm Tr}^{\prime}{\rm Tr}^{\prime \prime}e^{-\tilde{\mathcal H}_n}}=\frac{{\rm Tr}^{\prime}Ae^{-\tilde{\mathcal H}^{\prime}_n}}{{\rm Tr}^{\prime}e^{-\tilde{\mathcal H}^{\prime}_n}},
 \label{eq:deciA}
\end{equation}
where we have set $\tilde{\mathcal H}_n=\beta {\mathcal H}_n$ and defined
$\tilde{\mathcal H}^{\prime}_n$ by $\exp\left(-\tilde{\mathcal
H}^{\prime}_n\right)={\rm Tr}^{\prime \prime}\exp\left(-\tilde{\mathcal
H}_n\right)$. Note that the final expression can be regarded as the
thermal expectation value in the Ising model on the $(n-1)$-th
generation SG with the Hamiltonian $\tilde{\mathcal H}^{\prime}_n$. 

Because of the special geometry of the SG, the interactions in the new
Hamiltonian (as functions of the interactions in $\tilde{\mathcal
H}^{\prime}_n$) are independent of the generation $n$. Therefore, it
suffices to consider the case with $n=2$.

 \begin{figure}
\centerline{\epsfig{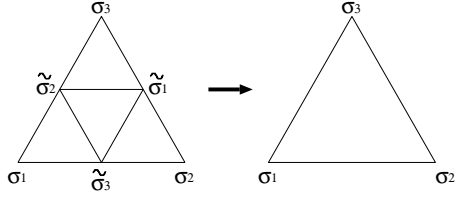}}
 \caption[dummy]{
 Thanks to the special geometry of the SG, the renormalization group map
 can be obtained if we consider a decimation procedure for the second generation
 SG.
 }
 	\label{fig:5}
 \end{figure}

Let ${\mathcal H}$ be a Hamiltonian on the second generation SG, and write
it as
\begin{eqnarray}
\tilde{\mathcal H}=\beta{\mathcal H} &=& -K_1( \tilde{\sigma}_{1}\tilde{\sigma}_{2}+\sigma_{2}\tilde{\sigma}_{3}+\tilde{\sigma}_{3}\sigma_{1} )  \nonumber\\
               & &-K_2( \tilde{\sigma}_{1}\sigma_{2}+\tilde{\sigma}_{2}\tilde{\sigma}_{3}+\sigma_{3}\tilde{\sigma}_{1} )  \nonumber\\
               & &-K_3( \sigma_{1}\tilde{\sigma}_{2}+\tilde{\sigma}_{2}\sigma_{3}+\tilde{\sigma}_{3}\tilde{\sigma}_{1}), 
\end{eqnarray} 
where $K_i \equiv \beta J_i, i=1,2,3$ (see Fig.~\ref{fig:5}). 
(Below, we use $K_i$ instead of $\beta J_i$.)
Because we have $K_i\propto1/T$, $T\to0$ implies $K_i \to
\infty$.
The renormalized parameters $K_1^{\prime}$, $K_2^{\prime}$ and $K_3^{\prime}$ are
determined from
\begin{equation}
\sum_{\tilde{\sigma}_{1} = \pm 1} \sum_{\tilde{\sigma}_{2} = \pm 1} \sum_{\tilde{\sigma}_{3} = \pm1} e^{-\beta {\mathcal H}} = e^{-\beta {\tilde{\mathcal H}}^{\prime}}
\label{eq:moto}
\end{equation}
and
\begin{equation}
 {\tilde{\mathcal H}}^{\prime}=-K_1^{\prime}\sigma_1\sigma_2-K_2^{\prime}\sigma_2\sigma_3-K_3^{\prime}\sigma_3\sigma_1.
\end{equation}
An explicit calculation shows that
\begin{eqnarray}
K_1^{\prime}= \frac{1}{4}\ln \left[\frac{B\cosh(2K_1)}{A\cosh(2K_2)\cosh(2K_3)}  \right], \label{eq:henkan1}\\
K_2^{\prime}= \frac{1}{4}\ln \left[\frac{B\cosh(2K_2)}{A\cosh(2K_3)\cosh(2K_1)}  \right], \label{eq:henkan2}\\
K_3^{\prime}= \frac{1}{4}\ln \left[\frac{B\cosh(2K_3)}{A\cosh(2K_1)\cosh(2K_2)}  \right], \label{eq:henkan3}
\end{eqnarray}
where
\begin{eqnarray}
A&=&\exp(K_1+K_2+K_3)+\sum_{i=1}^3 \exp(2K_i-K_1-K_2-K_3), 
\end{eqnarray}
and
\begin{eqnarray}
B&=&\exp(K_1+K_2+K_3)\cosh[2(K_1+K_2+K_3)]  \nonumber \\
 & &+\sum_{i=1}^3 \exp(2K_i-K_1-K_2-K_3)\cosh[2(K_1+K_2+K_3-2K_i)].
\end{eqnarray}

We set $K_i(0)=\beta J_i$ for $i=1,2,3$. We apply the renormalization
group (RG) map given by Eqs. (\ref{eq:henkan1})--(\ref{eq:henkan3}) to the initial
conditions $K_1(0)$, $ K_2(0)$ and $K_3(0)$ repeatedly. We denote the results of $n$ applications of the RG map by $K_1(n)$, $K_2(n)$ and $K_3(n)$.

\subsection{Correlation lengths}
Here we derive a general expression for the correlation lengths  using
the RG method. Let ${\mathcal H}_{n}$ be the Hamiltonian of the form
(\ref{eq:ham}) on the $n$-th generation SG. 
Then, using Eq. (\ref{eq:deciA}) $n$
times,
we find 
\begin{equation}
 \bkt{\sigma_{00}\sigma_{2^n 0}}^{(n+1)}=\langle \sigma_{00}\sigma_{10} \rangle_{\tilde{\mathcal H}_1}^{(1)},
 \label{eq:star}
\end{equation}
where the right-hand side represents the expectation value in the model
on the first generation SG, and
\begin{equation}
 \tilde{\mathcal H}_1^{\prime}=-K_1(n)\sigma_{00}\sigma_{01}-K_2(n)\sigma_{01}\sigma_{10}-K_3(n)\sigma_{10}\sigma_{00}.
\end{equation}
By explicitly calculating the right-hand side of Eq. (\ref{eq:star}), we find
\begin{equation}
 \langle \sigma_{00}\sigma_{10} \rangle_{\tilde{\mathcal H}_1}^{(1)}=\frac{e^{K_1(n)}\cosh\left(K_2(n)+K_3(n)\right)-e^{-K_1(n)}\cosh\left(K_2(n)-K_3(n)\right)}{e^{K_1(n)}\cosh\left(K_2(n)+K_3(n)\right)+e^{-K_1(n)}\cosh\left(K_2(n)-K_3(n)\right)} .\label{eq:ept}
\end{equation}
To evaluate the correlation lengths, we let $n_0$ be a sufficiently large
number such that $K_i(n)\ll 1$ holds for all $n\geq n_0$. Then, for
$n\geq n_0$, the RG map given by Eqs. (\ref{eq:henkan1})--(\ref{eq:henkan3}) is
approximated by
\begin{equation}
 K_i(n+1)\simeq K_i(n)^2,
\end{equation}
whose solution is
\begin{equation}
 K_i(n)\simeq K_i(n_0)^{2^{n-n_0}},
\end{equation}
for $i=1,2,3$. 

When quantities $K_i(n)$ are all sufficiently small, the right-hand side of
 Eq. (\ref{eq:ept}) is approximated by
$K_1(n)$.
Then, from the definition (\ref{eq:modoki}), we immediately get
\begin{equation}
 \xi_1(T)\simeq -\frac{2^{n_0}}{\ln K_1(n_0)}. \label{eq:xi1}
\end{equation} 
Exactly the same analysis is used for the other directions, and we
get
\begin{equation}
 \xi_i(T)\simeq -\frac{2^{n_0}}{\ln K_i(n_0)}
 \label{eq:xii}
\end{equation}
for $i=1,2$ and 3.

\subsection{Behavior of the correlation length in the isotropic case}
Before dealing with the problem of isotropy restoration, we briefly
examine the singular behavior of the correlation length in the isotropic
model with $J_1=J_2=J_3$. (Because $K_i(n)$ are identical for $i=1,2,3,$
we drop the subscript $i$ hereafter.) Since we have the expression
(\ref{eq:modoki}), it suffices to evaluate $n_0$.

\begin{figure}
\centerline{\epsfig{file=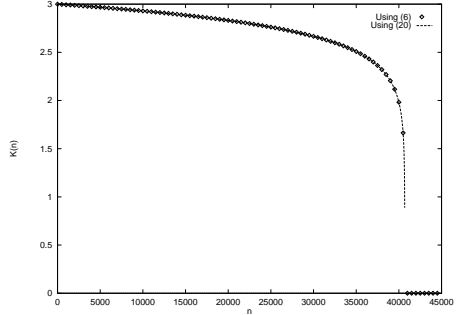,width=6cm}}
        \caption[dummy]{Each diamond denotes
          500 points of the renormalization flow,
         while the curve is the 
 approximate RG flow of Eq. (\ref{eq:kn}).}
        \label{fig:K(n)}
 \end{figure}

As we are interested  in the $T\to0$ limit, we assume $K(0)\gg1$ and
note that the quantities in Eqs. (\ref{eq:henkan1})--(\ref{eq:henkan3}) are approximated as
\begin{equation}
 K(n+1)\simeq K(n)-\exp(-4K(n))
 \label{eq:map}
\end{equation}
if $K(n)\gg 1$ holds. This difference relation is  approximated accurately
by the differential equation
\begin{equation}
 \frac{ d}{{ d}n}K(n)=-e^{4K(n)},
\end{equation}
whose solution is
\begin{equation}
 K(n)=\frac{1}{4}\ln\left(e^{4K(0)}-4n\right). \label{eq:kn}
\end{equation}
To check the accuracy of the approximation 
we compare in Fig.~\ref{fig:K(n)}
the RG flow obtained by numerically solving
Eqs. (\ref{eq:henkan1})--(\ref{eq:henkan3}) with the solution
(\ref{eq:kn}). We find that the approximation (\ref{eq:kn}) is quite
accurate as long as $K(n)\gtrsim 1$. Furthermore, we observe that $K(n)$
immediately enters the region $K(n)\ll1$ when the condition
$K(n)\gtrsim1$ is violated.
Based on the second observation, we assume that the behavior of the RG
flow in the regions $K(n)\ll1$ and $K(n)\gg1$ can be connected almost
smoothly. 

To determine $n_0$, we set $K(n_0)=K_0$, where
$K_0\ll1$ is an arbitrary constant that we shall fix.
Then, from Eq. (\ref{eq:kn}), we get
\begin{equation}
 n_0=\frac{1}{4}\left(e^{4\,K(0)}-e^{4\,K_0}\right).
\end{equation} 
By substituting this expression into Eq. (\ref{eq:xii}), we finally get
\begin{eqnarray}
 \xi(T)\simeq \frac{2^{\frac{1}{4}\left(e^{4K(0)}-e^{4K_0}\right)}}{\ln K_0}\propto \exp\left[{\rm const.}\,e^{4K(0)}\right]=\exp\left[C\,e^{4J/(k_BT)}\right],
\end{eqnarray}
in agreement with Ref. \cite{GMA}. Here, $C$ is a constant (which cannot be
determined precisely from the present crude analysis).

\subsection{Partial and perfect restoration of isotropy}
We now treat the general case with $K_1(0)\leq K_2(0)\leq K_3(0)$. We
concentrate on the region in which $K_i(n)\gg 1$.
The RG flow in this
region essentially determines the low temperature behavior of the
system. Here, the RG map (\ref{eq:henkan1}) is approximated by
\begin{eqnarray}
 K_1 (n+1) &\simeq& K_1 (n)-\frac{1}{4} \Big(\,e^{-2(K_1 (n)+K_2
 (n))}+e^{-2(K_2 (n)+K_3(n))} \nonumber \\
& &+e^{-2(K_3 (n)+K_1 (n))}-e^{-4K_1 (n)}+e^{-4K_2 (n)}+e^{-4K_3 (n)} \Big).  \label{eq:reno}
\end{eqnarray}
We get similar equations for $K_2$ and $K_3$ from Eqs. (\ref{eq:henkan2}) and
(\ref{eq:henkan3}), respectively. (They are obtained by cyclic
permutations of the indices 1, 2 and 3.)

\begin{figure}
\centerline{\epsfig{file=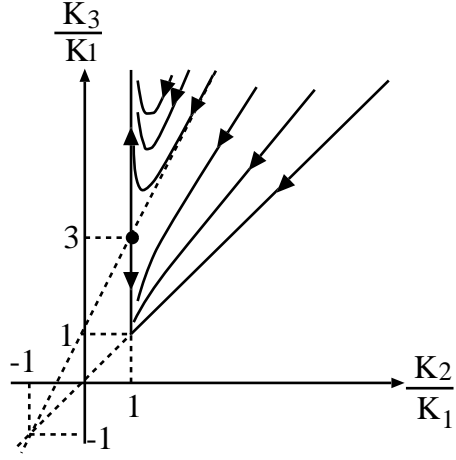,width=6cm}}
 	\caption{The RG flow in the two-dimensional space $(K^{(n)}_2/K^{(n)}_1,K^{(n)}_3/K^{(n)}_1)$ obtained from (\ref{eq:reno}).
There are three fixed points, $(K^{(n)}_2/K^{(n)}_1,K^{(n)}_3/K^{(n)}_1)=(1,\infty)$, $(1,3)$ and $(1,1)$.}
 	\label{fig:7}
 \end{figure}

Let us first focus on the behavior of the RG flow on the line
$K_2(n)/K_1(n)=1$ and further assume that $K_3(n)-K_1(n)\gg 1$. Then
the RG map (\ref{eq:reno}) is approximated by
\begin{eqnarray}
 \frac{K_3(n+1)}{K_1(n+1)}-\frac{K_3(n)}{K_1(n)} \simeq \left(4K_1(n)e^{4K_1(n)}\right)^{-1}\left(\frac{K_3(n)}{K_1(n)}-3\right).
 \label{32}
\end{eqnarray}
This implies that in the two-dimensional space, the point
$(K_2(n)/K_1(n),K_3(n)/K_1(n))=(1,3)$ is a fixed point. Moreover, we find
that there are RG flows which emerge from the fixed point, as depicted in Fig.~\ref{fig:7}.

Next we consider the case with $K_2(n)/K_1(n)>1$, which is the region 1
in Fig.~\ref{fig:4}. From Eq. (\ref{eq:reno})
we find that $K_2(n+1)/K_1(n+1)>K_2(n)/K_1(n)$ and the RG map
(\ref{eq:reno}) is approximated by 
\begin{equation}
\left({\frac{K_3(n)}{K_1(n)}+1}\right)
\left({\frac{K_2(n)}{K_1(n)}+1}\right)^{-1}
\simeq 
\left({\frac{K_3(n+1)}{K_1(n+1)}+1}\right)
\left({\frac{K_2(n+1)}{K_1(n+1)}+1}\right)^{-1}.  
\label{eq:sengai}
\end{equation}
It is hypothesized that the RG flow in the case with $K_2(n)/K_1(n)>1$ is
like that depicted in Fig.~\ref{fig:7}. In the case with $K_3(n)/K_1(n)\geq
2K_2(n)/K_1(n)+1$, the RG flow goes to the fixed point
$(K_2(n)/K_1(n), K_3(n)/K_1(n))=(1,\infty)$, and in the case with
$K_3(n)/K_1(n)< 2K_2(n)/K_1(n)+1$, the RG flow goes to the fixed point
$(K_2(n)/K_1(n), K_3(n)/K_1(n))=(1, 1)$.
We  confirm this picture in each case in what follows.

\subsubsection{Partial restoration of isotropy}
Now we treat the case with
$K_3(n)/K_1(n)\geq(2K_2(n))/K_1(n)+1$. We separately consider three
different regimes of $n$. 
Let us define $n_1$  by the relations $K_1(n_1)\simeq K_2(n_1)\simeq 1,K_3(n_1)\gg1$.
(Below we confirm that such an
$n_1$ exists.)
Then Eq. (\ref{eq:reno}) transforms into 
\begin{eqnarray}
\frac{dK_1(n)}{dn} &\simeq& -\frac{1}{4}\left(e^{-2(K_1 (n)+K_2 (n))}-e^{-4K_1(n)}+e^{-4K_2(n)}\right),  \label{eq:l1}\\
\frac{dK_2(n)}{dn} &\simeq& -\frac{1}{4}\left(e^{-2(K_1 (n)+K_2 (n))}+e^{-4K_1(n)}-e^{-4K_2(n)} \right), \label{eq:l2}\\
\frac{dK_3(n)}{dn} &\simeq& -\frac{1}{4}\left(e^{-2(K_1 (n)+K_2 (n))}+e^{-4K_1(n)}+e^{-4K_2(n)} \right),\label{eq:l3}
\end{eqnarray}
where we have used the relation $K_3(n)-K_1(n)\gg1$. These equations can be solved
analytically. Letting 
$L_1(n)=e^{-2(K_1(n)+K_2(n))}, L_2(n)=e^{-2(K_2(n)+K_3(n))}$ and $ L_3(n)=e^{-2(K_3(n)+K_1(n))}$, the equations (\ref{eq:l1}), (\ref{eq:l2}) and (\ref{eq:l3}) become 
\begin{eqnarray}
\frac{dL_1(n)}{n} &\simeq& L_1(n)^2,              \label{eq:L1}\\
\frac{dL_2(n)}{n} &\simeq& L_1(n)( L_2(n)+L_3(n))\label{eq:L2}
\end{eqnarray}
and
\begin{equation}
\frac{dL_3(n)}{n} \simeq L_1(n)( L_2(n)+L_3(n)),\label{eq:L3}
\end{equation}
respectively.
Then, from Eq. (\ref{eq:L1}), we have
\begin{equation}
L_1(n)\simeq\frac{L_1(0)}{1-L_1(0)n}.  
\end{equation}
Next, we define $L_\pm (n)\equiv L_2(n)\pm L_3(n)$, and we have
\begin{equation}
\frac{dL_+ (n)}{dn} \simeq 2L_1(n)L_+(n)=\frac{2L_1(0)}{1-L_1(0)n} L_+(n).
\end{equation}
Solving this differential equation, we get
\begin{equation}
L_+(n)\simeq \frac{L_2(0)+L_3(0)}{(1-L_1(0)n)^2}.\label{eq:l+}
\end{equation}
We can obtain $L_-(n)$ similarly, using
\begin{equation}
 \frac{dL_-(n)}{dn} \simeq 0 ,
\end{equation}
from which we have
\begin{equation}
L_-(n) \simeq  L_-(0)=L_2(0)-L_3(0).\label{eq:l-}
\end{equation}
From Eqs. (\ref{eq:l+}) and (\ref{eq:l-}), we find
\begin{eqnarray}
L_2(n)&=& \frac{L_+(n)+L_-(n)}{2} \nonumber\\
      &\simeq& \frac{L_2(0)+L_3(0)+(L_2(0)-L_3(0))(1-L_1(0)n)^2}{2(1-L_1(0)n)^2} , \\ 
L_3(n)&=& \frac{L_+(n)-L_-(n)}{2}  \nonumber\\
      &\simeq& \frac{L_2(0)+L_3(0)-(L_2(0)-L_3(0))(1-L_1(0)n)^2}{2(1-L_1(0)n)^2}.
\end{eqnarray}
Therefore $K_1(n),K_2(n)$ and $K_3(3)$ are finally expressed as
\begin{eqnarray}
K_1(n) &\simeq& \frac{1}{4}\ln\left[\frac{L_2(n)}{L_1(n)L_3(n)}\right],  \label{eq:l_1}\\    
K_2(n) &\simeq& \frac{1}{4}\ln\left[\frac{L_3(n)}{L_1(n)L_2(n)}\right],  \label{eq:l_2}\\    
K_3(n) &\simeq& \frac{1}{4}\ln\left[\frac{L_1(n)}{L_2(n)L_3(n)}\right].  \label{eq:l_3}    
\end{eqnarray}
From Eq. (\ref{eq:l_1}), we see that $n_1$ can be taken as
\begin{equation}
n_1\simeq e^{2(L_1(0)+L_2(0))}. \label{eq:n_1=}
\end{equation}
We can confirm that $K_1(n_1)\simeq K_2(n_1)\simeq 1$ by substituting $n_1$ into Eqs.
(\ref{eq:l_1}) and (\ref{eq:l_2}).

\begin{figure}
\centerline{\epsfig{file=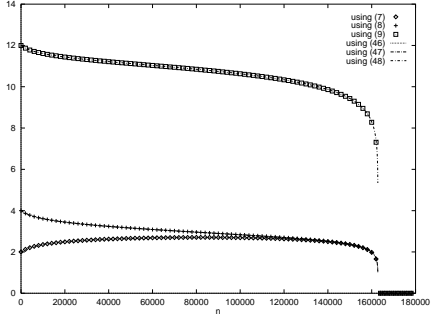,width=6cm}}
 	\caption[dummy]{Each square,  plus sign, and  diamond represents
 	 2000 points of $K_1$, $K_2$ and $K_3$, respectively.
 	The curves are the corresponding approximate RG flows of Eqs.
 	 (\ref{eq:l_1})--(\ref{eq:l_3}).}
 	\label{fig:8}
\end{figure}

In Fig.~\ref{fig:8}, we compare the approximate RG flows
 (\ref{eq:l_1})--(\ref{eq:l_3}) with numerical data.
 The agreement is again quite satisfactory.

Next we consider the regime in which $n_2>n>n_1$, in the case that $n_2$ is such that
$K_1(n_2)\ll1, K_2(n_2)\ll1$ and $K_3(n_2)\simeq1$ . In this regime Eqs.
(\ref{eq:henkan1}), (\ref{eq:henkan2}) and (\ref{eq:henkan3}) are
approximated as 
\begin{eqnarray}
K_1(n+1)&\simeq& 4K_1(n) \label{eq:al1},\\
K_2(n+1)&\simeq& 4K_2(n) \label{eq:al2},\\
K_3(n+1)&\simeq& K_3(n)-\frac{1}{2}\ln2.\label{eq:al3}
\end{eqnarray} 
The relations (\ref{eq:al1}), (\ref{eq:al2}) and (\ref{eq:al3}) are solved as
\begin{eqnarray}
K_1(n)&\simeq& (2^{n-n_1}-1)K_1(n_1)^{2^{n-n_1}}, \label{eq:k1=}\\
K_2(n)&\simeq& (2^{n-n_1}-1)K_2(n_1)^{2^{n-n_1}}, \label{eq:k2=}\\
K_3(n)&\simeq& K_3(n_1)-\frac{(n-n_1)}{2}\ln2 .\label{eq:k3=}
\end{eqnarray}
From Eq. (\ref{eq:k3=}) and $K_3(n_2)\simeq1$, $n_2$ is estimated as
\begin{equation}
n_2\simeq n_1+\frac{2}{\ln2}\left(K_3(0)-K_1(0)-2K_2(0)\right).
\end{equation}
The correlation lengths are estimated by using Eqs.
(\ref{eq:xii}) and (\ref{eq:k1=})--(\ref{eq:k3=}):
\begin{eqnarray}
\xi_1(T)&\simeq& \frac{1}{-\ln K_1(n_1)}2^{\exp[2(K_1(0)+K_2(0))]}\\
        &\sim& 2^{\exp[2(K_1(0)+K_2(0))]},\\
\xi_2(T)&\simeq& \frac{1}{-\ln K_2(n_1)}2^{\exp[2(K_1(0)+K_2(0))]}\\
        &\sim& 2^{\exp[2(K_1(0)+K_2(0))]},\\
\xi_3(T)&\simeq&\frac{1}{-\ln K_3(n_0)}2^{\exp[2(K_1(0)+K_2(0))]+\frac{2}{\ln2}\left(K_3(0)-K_1(0)-2K_2(0)\right) }\\
        &\sim& 2^{\exp[2(K_1(0)+K_2(0))]+\frac{2}{\ln2}\left(K_3(0)-K_1(0)-2K_2(0)\right) }   .
\end{eqnarray}
The ratio of $\xi_3(T)$ to $\xi_1(T)$ [or $\xi_2(T)$] diverges as
$T\rightarrow \infty$.

\subsubsection{Perfect restoration of isotropy}
\begin{figure}
\centerline{\epsfig{file=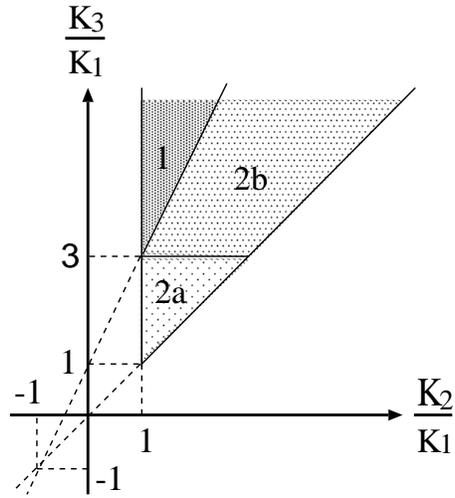,width=6cm}}
 	\caption{The region 2a with $K_2/K_1>1$, $K_3/K_1<3$, and the region 2b with $K_3/K_1<2(K_2/K_1)+1$, $K_3/K_1\ge3$.}
 	\label{fig:10}
\end{figure}

Finally, we show that in the region 2, we have  $K_2/K_1\rightarrow 1$ and
$K_3/K_1\rightarrow 1$ as $T\rightarrow0$; i.e, there is perfect
restoration of  isotropy. We treat the following three cases
separately.

\bigskip\noindent
\textit{(i) The case  $K_2/K_1=1$.}

\begin{figure}
\centerline{\epsfig{file=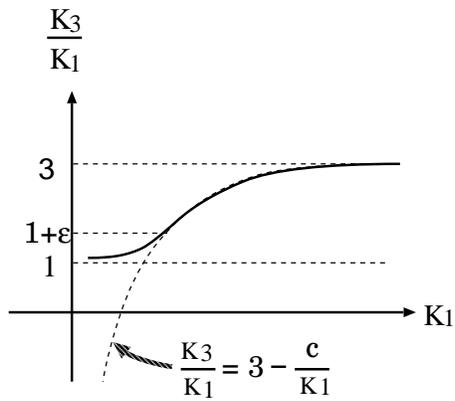,width=6cm}}
 	\caption{The RG flow in the two-dimensional space $(K_1(n),K_3(n)/K_1(n))$.}
 	\label{fig:11}
\end{figure}

Assuming $K_3(n)-K_1(n)\gg1$ in the RG map (\ref{eq:reno}), we get
\begin{equation}
\frac{K_3(n)}{K_1(n)}-\frac{K_3(n+1)}{K_1(n+1)}\simeq \frac{e^{4K_1(n)}}{4K_1(n)}\left\{ 3-\frac{K_3(n)}{K_1(n)} \right\}
\label{eq:star1}
\end{equation}
and
\begin{equation}
K_1(n)-K_1(n+1)\simeq\frac{e^{-4K_1(n)}}{4}.\label{eq:star2}
\end{equation}
Then, with $x_n\equiv K_1(n)$ and $y_n\equiv K_3(n)/K_1(n)$, these two relations yield
\begin{equation}
\frac{y_{n+1}-y_n}{x_{n+1}-x_n}\simeq\frac{3-y_n}{x_n}.
\end{equation}
This becomes the differential equation $dy/dx=(3-y)/x$ whose solution is
$y=3-C/x$,  with an arbitrary constant $C$. Therefore, $K_1(n)$ and
$K_3(n)/K_1(n)$ vary on the curve
\begin{equation}
\frac{K_3(n)}{K_1(n)}=3-\frac{C}{K_1(n)}\label{eq:curve}
\end{equation}
as in  Fig.~\ref{fig:11}.
 When $n$ is large and the assumption $K_3(n)-K_1(n)\gg1$ is no longer
 valid, the RG flow deviates from the curve (\ref{eq:curve}). Let
 $1+\epsilon$ be the value of $K_3(n)/K_1(n)$ at the value of $n$ where this deviation starts
 to take place. We can take $\epsilon$ as small as we wish by letting
 $T\rightarrow 0$ (i.e., $K_1(0)\rightarrow\infty$).
 Therefore, we conclude that $K_3/K_1\to1$ as $T\to0$.

\bigskip\noindent
\textit{(ii) The cases  $K_2/K_1>1$ and $K_3/K_1<3$ (the region 2a in Fig.~\ref{fig:10}).}

 As in the case (i), the RG map (\ref{eq:reno}) and the assumption
 $K_3(n)-K_1(n)\gg1$ yield the equation (\ref{eq:star1}). The equation
 (\ref{eq:star2}) is modified as
\begin{equation}
K_1(n)-K_1(n+1)\lesssim\frac{e^{-4K_1(n)}}{4},
\end{equation}
which is a stronger relation than Eq. (\ref{eq:star2}). Therefore, the same
argument as in (i) shows that $K_3(n)/K_1(n)$ can be made arbitrarily
close to $1$ by letting $T\rightarrow 0$.

\bigskip\noindent
\textit{(iii) The case  $K_3/K_1<2(K_2/K_1)+1$ and $K_3/K_1\geq3$
(the region 2b in Fig.~\ref{fig:10}).} 

From Eq. (\ref{32}), we find that the RG flow depicted in the
$(K_2/K_1,K_3/K_1)$ plane follows the curves in Fig.~\ref{fig:10}. Thus the RG
trajectories flow from the region 2b into the region 2a. In what
follows, we show that the conditions $K_3(n)-K_1(n)\gg1$ and
$K_2(n)-K_1(n)\gg1$ are satisfied when the trajectory enters the region
2a, provided that we take a sufficiently small values of  $T$. Then we can  use
 the argument given in (ii) to conclude that a perfect 
restoration of isotropy takes place for $T\rightarrow0$.
Supposing that $K_3(n)-K_1(n)\gg1$ and $K_2(n)-K_1(n)\gg1$, the RG map
(\ref{eq:reno}) yields
\begin{equation}
K_1(n)-K_1(n+1)\simeq-\frac{e^{-4K_1(n)}}{4}.\label{eq:star4}
\end{equation}
Because this shows that $K_1(n)$  increases with $n$, we see that the
desired condition $K_3(n)-K_1(n)\gg1$ is valid when the trajectory
enters the region 2a.
Let $a$ be the value of $K_2(n)/K_1(n)$ at which the trajectory enters 2a. By
taking $K_1(0)$ so that $(a-1)K_1(0)\gg1$, we can guarantee that the
desired condition $K_2(n)-K_1(n)\gg1$ is valid when the trajectory
enters 2a. 

\subsection{The differences between the present and previous studies}
\label{diff}
Brody and Ritz \cite{BR}, and later Bhattacharyya and Chakrabarti \cite{BC} studied the same problem of the restoration of isotropy in the Ising model on the Sierpi\'nski gasket.
They both started from a partially anisotropic model with $J_2=J_3$ or $J_1=J_2$, and studied the model using the same RG equation.
But the conclusion of  Ref. \cite{BR} is that there is complete isotropy restoration while that of Ref.  \cite{BC} is that isotropy restoration is impossible.
Note that our conclusion that the restoration may be  either partial or perfect agrees with neither of these two conclusions.
We stress that our analysis is much more elaborate and careful than the previous two, and it therefore provides the most reliable conclusion.

Let us point out some of the inaccurate points in the previous works, which may be the sources of (partially) incorrect conclusions.

In Ref. \cite{BC}, the authors make a rather crude ``weak coupling'' approximation to the RG map, which effectively brings the system to an extremely high temperature by a small number of application of the RG map.
Thus is is not possible for nonlinear effects to exist, and hence there results no isotropy restoration.

In Ref. \cite{BR}, the authors make an unjustified approximation using a philosophy very different from that in Ref. \cite{BC}.
When deriving their key equation, (7), the authors rely on a ``rescaling'' argument which can never be exact.
Consequently, (7) becomes an approximation which is probably valid only when $K_{n+1}\sim1$.
But this condition cannot be assumed for $L_n\ll1$.

In fact, our analysis clearly demonstrates that the present problem cannot be understood using a single approximate RG map.
Instead, one needs to carefully use different approximations depending on the ranges of the parameters, as we have done here.

\bigskip\noindent
{\bf Acknowledgements:}
I wish to thank Professor Hisao Hayakawa for useful discussions and continual encouragement.
Part of the present work was done when the author was at Gakushuin University.
I wish to thank Professor Hal Tasaki for his suggestion of the present problem and useful discussions.
I  thank Hiroto Kuninaka and Kim Hyeon-Deuk for kindly assisting me in the preparation of the paper.
I also thank an anonymous referee for letting me know of crucial references.

\end{document}